\documentclass[11pt]{article}
\usepackage[margin=1in]{geometry}
\usepackage{graphicx}
\usepackage{float}
\usepackage{caption}
\usepackage{subcaption}
\usepackage{amsmath}
\usepackage{booktabs}
\usepackage[space]{grffile}
\usepackage{url}
\usepackage{titling}
\providecommand{\tightlist}{%
  \setlength{\itemsep}{0pt}\setlength{\parskip}{0pt}}
\graphicspath{{./}}

\preauthor{\begin{center}\large}
\postauthor{\\[-0em]\normalsize\textit{Polar Dynamix}\par\end{center}}
\title{Digital Twin Assessment of Filter Clogging Penalties in VFD-Driven Industrial Fan Systems}
\author{Wichai Pattanapol and Adam Gill}
\date{2026-01-17}
\begin{document}
\maketitle
\begin{abstract}
Industrial ventilation systems equipped with variable-frequency drives (VFDs) often mask the aerodynamic impact of filter clogging by automatically increasing fan speed to maintain airflow setpoints. While effective for process stability, this control strategy creates a ``blind spot" in energy management, leading to unmonitored power spikes. This study applies a rapid digital-twin workflow to quantify these hidden energy penalties in a standard 50 kW draw-through fan room. Using a specialized computational fluid dynamics (CFD) solver (AirSketcher), the facility was modeled under ``Clean Filter" (baseline) and ``Dirty Filter" (clogged) scenarios. The physics engine was first validated against wind tunnel experimental data, confirming high agreement with the theoretical inertial pressure-drop law ($\Delta P \propto U^2$).
In the industrial case study, results indicate that severe clogging (modeled via a 50\% effective porosity reduction) can push the fan system beyond its available pressure head or speed limits, forcing the VFD into a saturation regime. Under these conditions, effective airflow collapses by over 50\% (3,806 CFM to 1,831 CFM) despite increased fan effort. The associated energy analysis predicts an annual energy penalty of 8,818 kWh (\$1,058/yr). This study demonstrates how a physics-based simulation provides a defensible, ROI-driven metric for optimizing filter maintenance cycles.
\end{abstract}
\section{Introduction}
\label{introduction}
In modern industrial facilities, air handling units (AHUs) are increasingly equipped with variable-frequency drives (VFDs) to maintain constant volumetric airflow (CFM) despite fluctuating system conditions. While beneficial for maintaining consistent environmental parameters, this control strategy introduces a latent failure mode: as filter media accumulates particulate matter, the VFD masks the problem by ramping up motor RPM to overcome the added resistance [1]. Unlike fixed-speed systems where clogging results in a noticeable drop in airflow, VFD-driven systems maintain ventilation performance but at a significantly higher energy cost, often going undetected until the motor approaches its operational limits.
When inlet resistance becomes sufficiently severe, the fan may reach either its maximum speed or available pressure head, at which point airflow can no longer be maintained. In this saturation regime, both energy consumption and ventilation performance are adversely affected.
This dynamic is particularly critical in precision manufacturing sectors. A prime example is \textbf{glass container manufacturing}, where large-capacity blowers provide essential cooling air for furnaces and bottle-forming machines. In this high-stakes environment, precise airflow control is not merely an energy efficiency concern but a strict quality mandate. A deviation in CFM caused by uncompensated filter impedance can alter cooling rates during the forming process, leading to thermal stress fractures, structural defects, and ultimately, entire batches of rejected products [2].
Quantifying this energy penalty requires correlating the aerodynamic resistance of the filter bank directly to the fan's power consumption. Classical fan laws dictate that power scales with the cube of the speed ($P \propto n^3$) or the product of flow and pressure drop ($P \propto Q \cdot \Delta P$). However, accurately calculating $\Delta P$ for complex, non-uniform filter geometries often relies on oversimplified empirical curves that fail to account for specific room dynamics [3].
This study utilizes a 2D computational fluid dynamics (CFD) tool, AirSketcher, to construct a "digital twin" of a factory fan room. The objective is to utilize a physics-based simulation to: (1) \textbf{validate} the porous media drag model against published wind tunnel data [3], and (2) \textbf{simulate} the operational and financial impact of a clogged filter in a draw-through industrial configuration.
\section{Methods}
\label{methods}
\subsection{Computational Framework}
The simulations were performed using \textbf{AirSketcher} [4], a steady-state incompressible Reynolds-Averaged Navier-Stokes (RANS) solver. The tool employs a structured Cartesian grid with automated local refinement to resolve sharp velocity gradients near filter faces without manual meshing overhead.
\subsubsection{Governing Equations}
The flow field is resolved via the incompressible Navier-Stokes equations for mass and momentum conservation:
\[ \nabla \cdot \mathbf{V} = 0 \]
\[ \rho \left( \frac{\partial \mathbf{V}}{\partial t} + \mathbf{V} \cdot \nabla \mathbf{V} \right) = -\nabla p + \mu \nabla^2 \mathbf{V} + \mathbf{S} \]
Where $\rho$ is fluid density, $\mu$ is dynamic viscosity, and $\mathbf{S}$ represents momentum source terms. Pressure-velocity coupling is enforced via a Poisson equation derived from the continuity constraint. Turbulence is modeled using the \textbf{Spalart-Allmaras (SA)} one-equation model [5], chosen for its robustness in bounded aerodynamic flows and ability to handle adverse pressure gradients effectively.
\subsubsection{Porous Media Formulation}
To replicate the aerodynamic resistance of industrial filter banks, a specialized drag formulation connects pressure drop to an effective porosity fraction $\phi$. While real-world filter fouling is spatially non-uniform, porosity reduction is employed here as a numerical proxy to represent the dominant increase in inertial resistance associated with particulate loading.
Numerical stability at low porosities is achieved via a regularized effective drag coefficient $C_d$:
\[ C_d = \frac{(1 - \phi)^2}{\phi^3 + \epsilon} \]
Within the porous zones, a semi-implicit damping term is applied to the velocity field to enforce the Darcy-Forchheimer resistance law.
\subsection{Validation Case: Porous Screen Experiment}
\begin{figure}[!ht]
    \centering
    \includegraphics[width=0.6\linewidth]{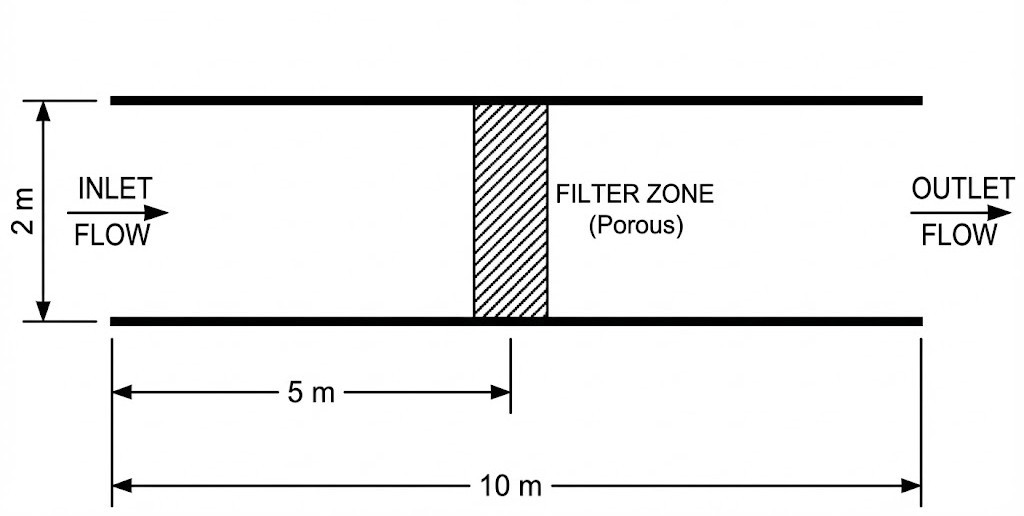}
    \caption{Wind Tunnel Simulation Layout (2D)}
    \label{fig:layout_tunnel}
\end{figure}

To ensure that the digital twin correctly predicts pressure drops across filter media, the solver was benchmarked against a wind tunnel experiment that was performed at Politecnico di Milano [3].
\begin{itemize}
\tightlist
    \item \textbf{Experimental Setup:} A virtual wind tunnel ($10 \text{ m} \times 2\text{ m}$) was configured with a central porous zone (Figure \ref{fig:layout_tunnel}).
    \item \textbf{Measurement:} The pressure drop was extracted using a \textbf{centerline probe} extending from the inlet to the outlet.
\end{itemize}
\subsection{Industrial Case Study: Factory Fan Room}
\begin{figure}[!ht]
    \centering
    \includegraphics[width=0.6\linewidth]{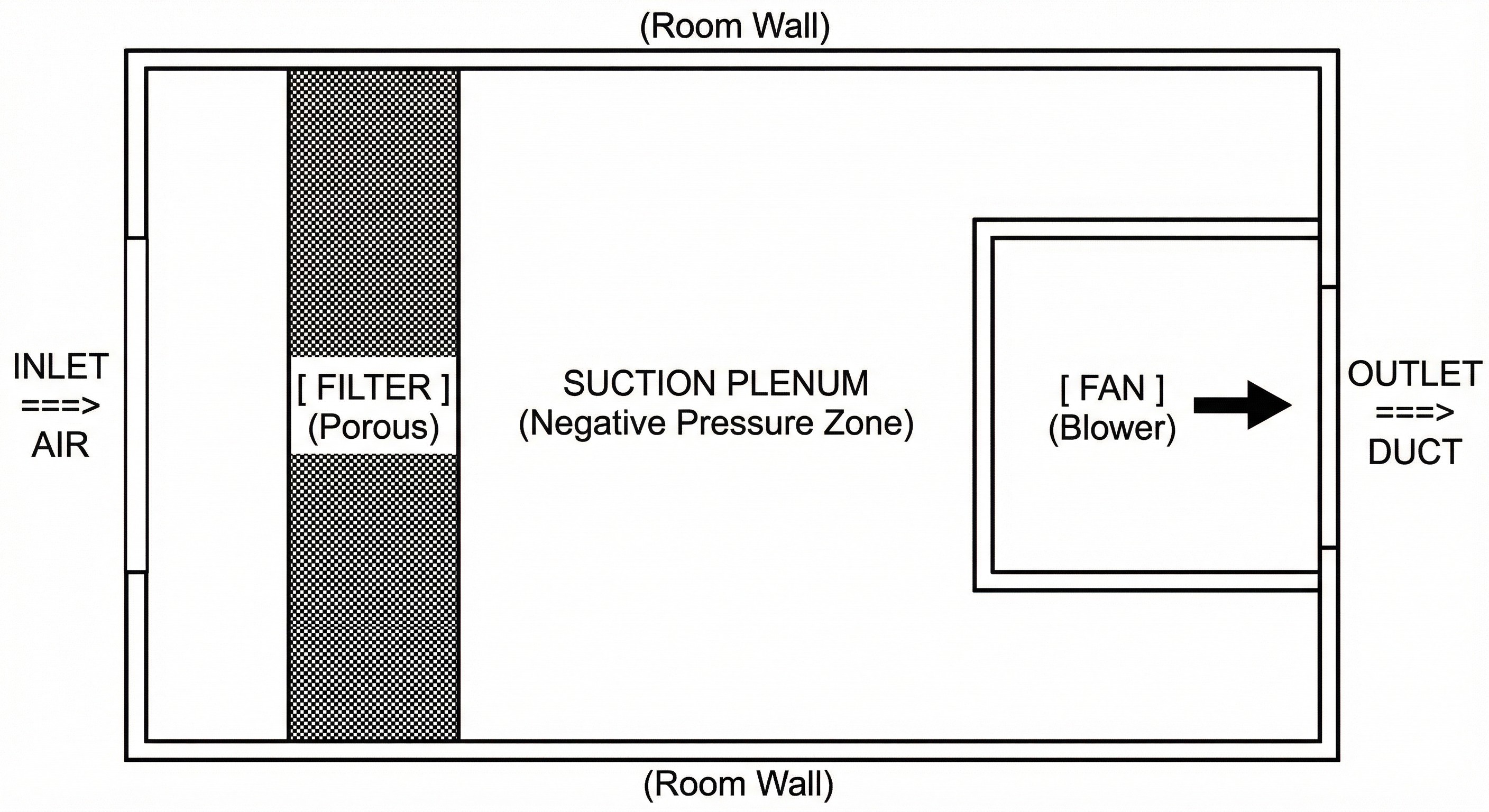}
    \caption{Factory Fan Room Layout (Plan View)}
    \label{fig:layout_fan}
\end{figure}

Following validation, the workflow was applied to a representative factory fan room (Figure \ref{fig:layout_fan}).
\begin{itemize}
\tightlist
    \item \textbf{Geometry:} A "draw-through" configuration (inlet $\rightarrow$ filter bank $\rightarrow$ suction plenum $\rightarrow$ fan) was applied.
    \item \textbf{Dimensions:} The room dimensions were 6 m (length) $\times$ width 4 m (width).
    \item \textbf{Scenarios:}
    \begin{enumerate}
    \tightlist
        \item \textbf{Clean filter (baseline):} The filter porosity was calibrated to low resistance (simulation setting: "80\% porosity").
        \item \textbf{Dirty filter (clogged):} Resistance was increased to simulate heavy loading (simulation setting: "50\% porosity").
    \end{enumerate}
    \item \textbf{Data extraction:} Performance metrics were sampled along a \textbf{longitudinal centreline probe} to capture velocity and pressure profiles upstream of the fan intake.
\end{itemize}
\section{Results}
\label{results}
\subsection{Validation Against Wind Tunnel Data}
\begin{figure}[!ht]
    \centering
    \includegraphics[width=0.9\linewidth]{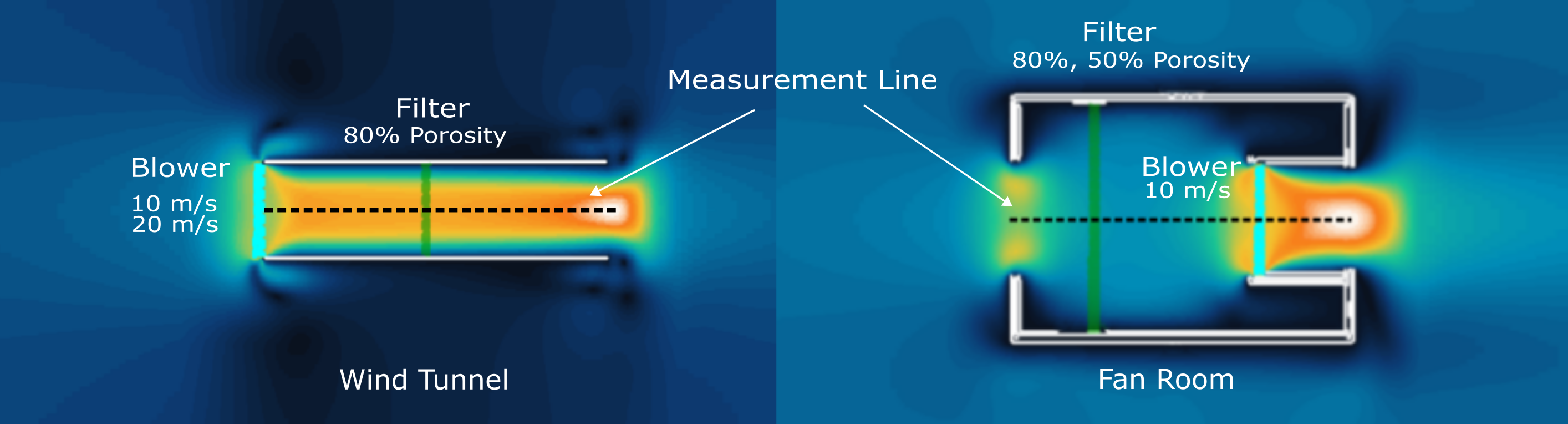}
    \caption{Location of the Centerline Measurement Probe}
    \label{fig:lineprobe}
\end{figure}

The calibration phase confirmed that the physics engine accurately reproduces the inertial resistance of porous screens.
\begin{table}[!ht]
\centering
\caption{Experimental Benchmark Data (Reference vs. Simulation)}
\label{tab:benchmark}
\begin{tabular}{lccc}
\toprule
\textbf{Inlet Velocity} ($U_{inlet}$) & \textbf{Exp. Target} ($\Delta P_{exp}$) & \textbf{AirSketcher} ($\Delta P_{sim}$) & \textbf{Deviation} \\
\midrule
10 m/s & 153 Pa & 165 Pa & +7.8\% \\
20 m/s & 612 Pa & 650 Pa & +6.2\% \\
\bottomrule
\end{tabular}
\end{table}
To ensure consistent data extraction, all velocity and pressure profiles presented in this study were sampled along a \textbf{longitudinal centerline probe} (Figure \ref{fig:lineprobe}).
\begin{figure}[!ht]
    \centering
    \begin{subfigure}{1\textwidth}
        \centering
        \includegraphics[width=\linewidth]{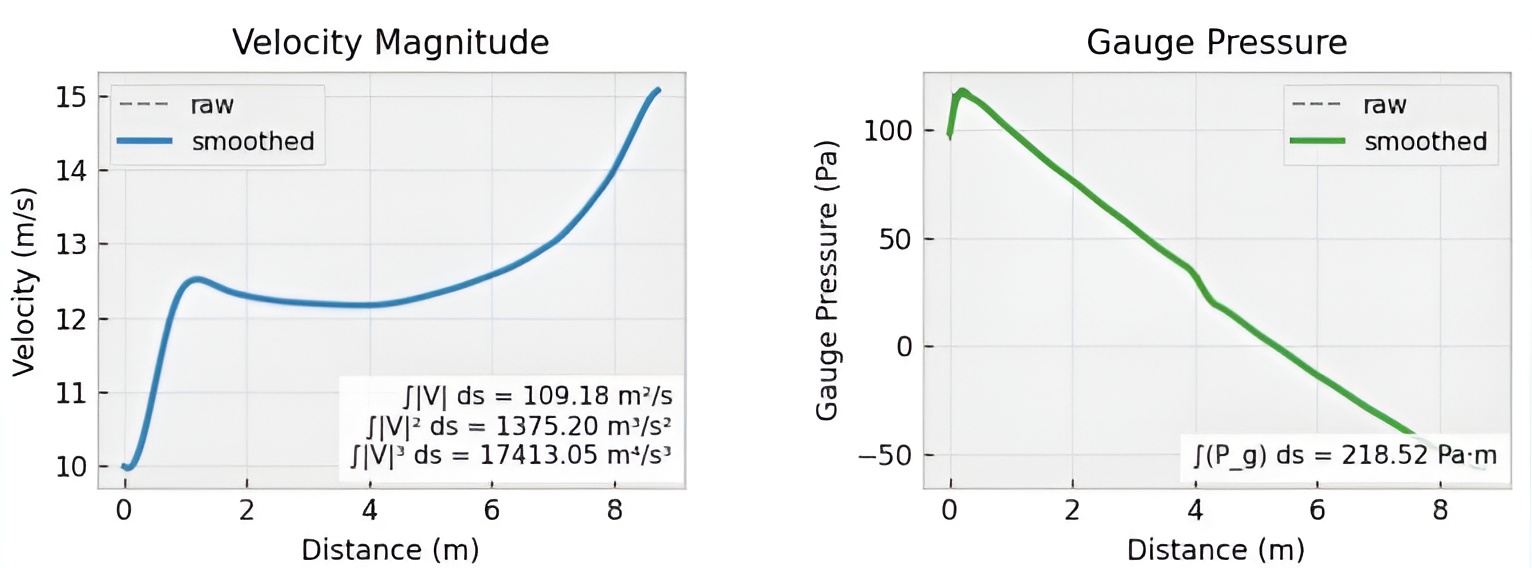}
        \caption{10 m/s Calibration}
        \label{fig:validation_10}
    \end{subfigure}
   
    \par\bigskip
    \begin{subfigure}{1\textwidth}
        \centering
        \includegraphics[width=\linewidth]{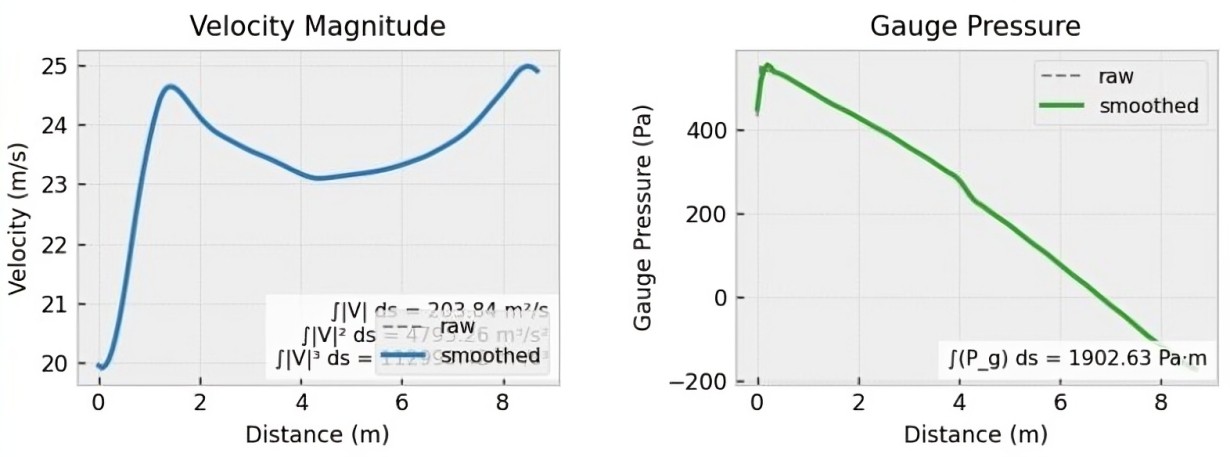}
        \caption{20 m/s Blind Test}
        \label{fig:validation_20}
    \end{subfigure}
    \caption{Validation of the digital twin physics engine against wind tunnel data.
    \textbf{(a)} 10 m/s (top): The upper plot displays the velocity magnitude (left) and gauge pressure (right) profiles at the baseline inlet velocity of 10 m/s.
    \textbf{(b)} 20 m/s (bottom): The lower plot displays the corresponding profiles at the doubled inlet velocity of 20 m/s, which was used to verify the quadratic pressure response. }
    \label{fig:validation_results}
\end{figure}
\subsection{Factory Fan Room: Clean Filter vs. Dirty Filter Comparison}
The validated model was applied to the factory geometry to quantify the aerodynamic impact of filter clogging.
\begin{figure}[!ht]
    \centering
    \begin{subfigure}{0.45\textwidth}
        \centering
        \includegraphics[width=\linewidth]{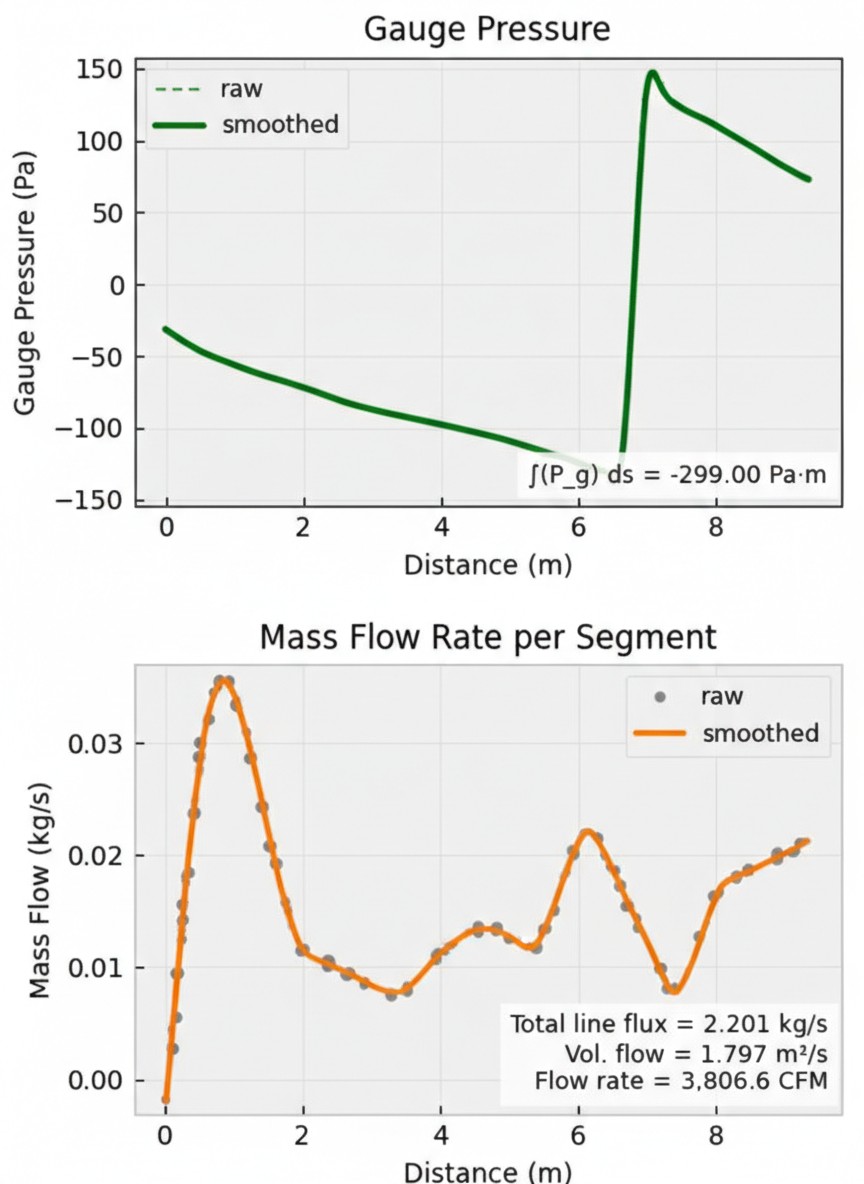}
        \caption{Clean Filter (3,806 CFM)}
        \label{fig:clean}
    \end{subfigure}
    \hfill
    \begin{subfigure}{0.45\textwidth}
        \centering
        \includegraphics[width=\linewidth]{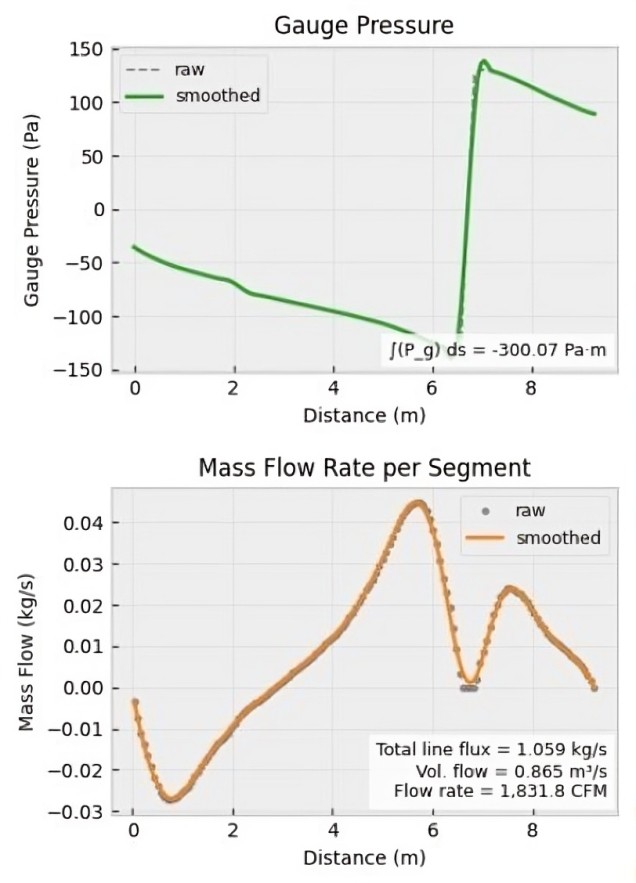}
        \caption{Dirty Filter (1,831 CFM)}
        \label{fig:dirty}
    \end{subfigure}
    \caption{Aerodynamic performance profiles sampled along the longitudinal centerline probe for the two simulation scenarios.
    \textbf{(a) Clean filter case (left):} The upper plot displays the gauge pressure distribution; the lower plot displays the mass and volumetric flow rate.
    \textbf{(b) Dirty filter case (right):} The upper plot displays the gauge pressure distribution; the lower plot displays the mass and volumetric flow rate.}
\end{figure}
Figures \ref{fig:clean} and \ref{fig:dirty} provide a comparative analysis of the system's hydraulic performance under the two maintenance conditions:
\begin{itemize}
\tightlist
    \item \textbf{Clean Filter Case (Figure \ref{fig:clean}):} Under baseline conditions (80\% porosity), the system exhibits a hydraulically efficient flow regime. The centerline velocity profile (left) is characterized by high, uniform magnitude, corresponding to a volumetric flow rate of \textbf{3,806 CFM}. The static pressure profile (right) indicates minimal impedance across the filter bank, permitting the fan to operate near its peak efficiency point on the system curve.
    \item \textbf{Dirty Filter Case (Figure \ref{fig:dirty}):} In the compromised state (50\% porosity), the flow physics are fundamentally altered. The effective volumetric flow rate collapses to \textbf{1,831 CFM}, representing a critical \textbf{52\% reduction} in ventilation capacity. The pressure signature (right) reveals a severe static vacuum spike immediately downstream of the filter face. This behavior represents a non-compressible choking-like condition driven by excessive inlet resistance, in which VFD energy is largely dissipated in generating suction head rather than inducing useful mass transport.
\end{itemize}
\section{Economic and Carbon Impact}
\label{economic-and-carbon-impact}
To translate these aerodynamic findings into actionable financial metrics, an energy analysis was performed using the simulation's automated reporting module.
\subsection{Calculation Methodology}
The values presented in Table 2 are derived directly from the \textbf{Upstream Energy Saving} summary in the simulation report:
\begin{enumerate}
\tightlist
    \item \textbf{Project inputs:}
    \begin{itemize}
    \tightlist
        \item \textbf{Baseline fan power:} 50.00 kW (contract/rated power)
        \item \textbf{Operation schedule:} 3,000 hours/year
        \item \textbf{Electricity tariff:} \$0.12 /kWh
    \end{itemize}
    \item \textbf{Energy calculation logic (upstream focus):}
    In a draw-through configuration, fan power consumption is driven by the work required to overcome inlet restriction. Fan shaft power is proportional to the product of volumetric flow and pressure rise ($P \propto Q \cdot \Delta P / \eta$). Assuming fan efficiency and downstream system resistance remain unchanged between cases, the digital twin estimates the energetic impact of inlet restriction using a combined proxy ratio ($r_{fan}$) derived from upstream flow and pressure measurements:
    \[ r_{fan} \approx r_Q \times r_{DP} = 0.987 \times 0.954 \approx 0.941 \]
    \begin{itemize}
    \tightlist
        \item \textbf{Net energy saving ($\Delta E$):} The model calculates the annual energy penalty that can be avoided by restoring the system to the "Clean Filter" baseline:
        \[ \Delta E = \text{Base Energy} \times (1 - r_{fan}) \]
        \[ (50 \text{ kW} \times 3000 \text{ h}) \times (1 - 0.941) \approx \textbf{8,818 kWh} \]
    \end{itemize}
    \item \textbf{Financial return (payback period):}
    Using the upstream-calculated energy delta and a conservative filter bank replacement cost estimate of \textbf{\$800} (assuming a bank of 10-12 standard industrial bag filters):
    \[ \text{annual cost saving} = 8,818 \text{ kWh} \times \$0.12 \approx \textbf{\$1,058} \]
    \[ \text{Payback Period} = \frac{\text{Maintenance Cost}}{\text{Annual Saving}} = \frac{\$800}{\$1,058} \approx \textbf{0.75 years (9 months)} \]
    \textit{Interpretation:} The maintenance investment pays for itself in under a year, significantly outperforming capital hardware retrofits.
\end{enumerate}
\begin{table}[!ht]
\centering
\caption{Annual Impact of Filter Replacement (Clean Filter vs. Dirty Filter Baseline)}
\begin{tabular}{lll}
\toprule
\textbf{Metric} & \textbf{Value} & \textbf{Implications} \\
\midrule
$\Delta$ Airflow & +1,975 CFM & Restoration of ventilation standards. \\
$\Delta$ Fan Energy & -8,818 kWh/yr & Energy saved by removing upstream restriction. \\
$\Delta$ Cost & -\$1,058 / yr & Direct financial saving (Payback: $<$ 9 mo). \\
$\Delta$ Carbon & -3.53 tCO2e & ESG contribution (Factor: 0.400). \\
\bottomrule
\end{tabular}
\end{table}
\section{Discussion}
\label{discussion}
The results of this study provide critical insights when benchmarked against broader industry data.
\textbf{Comparison with industry benchmarks:}
The simulation predicts an annual saving of \textbf{8,818 kWh} per 50 kW fan unit. This aligns with industry reports from the glass sector, where large-scale VFD retrofits on furnace air blowers have demonstrated savings of over \textbf{800,000 kWh/year} for facility-wide systems [2]. While our single-unit values are naturally lower, the physics remains consistent: overcoming flow resistance is a primary energy sink. Furthermore, the simulation shows a \textbf{52\% performance swing} in airflow due to clogging. This falls squarely within the \textbf{7--60\%} variance range observed in industrial VFD applications [2], confirming that fan systems are highly sensitive to inlet impedance.
\textbf{Scaling to factory operations:}
The impact of these findings scales significantly when applied to a full production line. A typical glass manufacturing facility may operate 10 to 20 similar large-capacity blowers for cooling and furnace applications. Extrapolating the single-unit penalty of \$1,058/year suggests a cumulative hidden cost of over \textbf{\$10,000 to \$20,000 annually} for a fleet of 10-20 fans. This validates the "Digital Twin" approach as a scalable tool for uncovering major operational inefficiencies.
\textbf{ROI: maintenance vs. retrofit:}
A key differentiator of this study is the return on investment (ROI). Industry standards for installing new VFD hardware typically cite a payback period of \textbf{1.7 years} [2]. In contrast, using simulations to trigger condition-based maintenance (assuming a realistic \$800 filter bank replacement cost) yields a payback period of approximately \textbf{9 months}. This highlights that while VFD hardware is essential, its efficiency is maximized only through rigorous aerodynamic maintenance.
\section{Conclusions}
\label{conclusions}
This study employed a digital-twin approach to analyze the operational and financial consequences of filter clogging in VFD-driven industrial fan systems. Based on the validation against wind tunnel data and the subsequent industrial case study, the following conclusions are drawn:
\begin{enumerate}
\tightlist
    \item \textbf{Physics fidelity:} The simulation framework successfully replicated wind tunnel pressure-drop data for porous media with a deviation of less than 8\%. This confirms the solver's reliability in predicting inertial resistance ($\Delta P \propto U^2$), which is a critical factor for modeling high-velocity industrial airflows.
    \item \textbf{System performance and quality risk:} The simulation revealed that a 50\% reduction in filter porosity resulted in a critical loss of ventilation performance, reducing airflow by \textbf{52\%} (from 3,806 to 1,831 CFM). In manufacturing contexts such as glass container production, such a deficit is not merely an efficiency loss but a severe operational risk that could compromise product quality and cooling rates.
    \item \textbf{Economic justification:} The energy analysis quantified a substantial annual penalty of \textbf{8,818 kWh} and \textbf{\$1,058} associated with the clogged state. This translates to a payback period of approximately \textbf{9 months} for filter replacement. This data provides facility managers with a clear, physics-backed business case for shifting from fixed-schedule maintenance to condition-based strategies, ensuring optimal performance while minimizing the facility's carbon footprint.
\end{enumerate}

\end{document}